\documentclass[lettersize,journal]{IEEEtran}
\usepackage{amsmath,amsfonts}
\usepackage{algorithmic}
\usepackage{amssymb}

\usepackage{algorithm}
\usepackage{array}
\usepackage[caption=false,font=normalsize,labelfont=sf,textfont=sf]{subfig}
\usepackage{textcomp}
\usepackage{stfloats}
\usepackage{url}
\usepackage{verbatim}
\usepackage{graphicx}
\usepackage{cite}
\usepackage{braket}
\DeclareUnicodeCharacter{2212}{\ensuremath{-}}

\begin{document}

\title{Efficient quantum non-fungible tokens for blockchain}

\author{Subhash Shankar Pandey, Tadasha Dash, Prasanta K. Panigrahi, and Ahmed Farouk
\thanks{ Subhash Shankar Pandey is with the Indian Institute of Science Education and Research Kolkata, Mohanpur 741246, West Bengal, India \\
E-mail: subhashshankarpandeyy@gmail.com 
}
\thanks{Tadasha Dash is with the Indian Institute of Science Education and Research Kolkata, Mohanpur 741246, West Bengal, India \\
E-mail: tadashadash08@gmail.com 
}
\thanks{ P. K. Panigrahi is with the Indian Institute of Science Education and Research Kolkata, Mohanpur 741246, West Bengal, India \\
E-mail: pprasanta@iiserkol.ac.in 
}
\thanks{ A. Farouk is with Department of Physics and Computer Science, Faculty of Science, Wilfrid Laurier University, Waterloo, Canada.\\ 
E-mail: afarouk@wlu.ca 
}
}

%The paper headers
%\markboth{ P. K. Panigrahi is with the Indian Institute of Science Education and Research Kolkata, Mohanpur 741246, West Bengal, India 
%E-mail: pprasanta@iiserkol.ac.in 
%}%
%{Shell %\MakeLowercase{\textit{et al.}}: A Sample Article Using IEEEtran.cls for IEEE Journals}


\maketitle

\begin{abstract}
Blockchain is a decentralized system that allows transaction transmission and storage according to the roles of the Consensus algorithm and Smart contracts. Non-fungible tokens (NFTs) consolidate the best characteristics of blockchain technology to deliver unique and bona fide tokens, each with distinctive attributes with non-fungible resources. Unfortunately, current classical NFTs are suffering from high costs regarding the consumed power of mining and lack of security. Therefore, this paper presents a new protocol for preparing quantum non-fungible tokens where a quantum state representing NFT is mounted on a blockchain instead of physically giving it to the owner. The proposed scheme is simulated and analyzed against various attacks and proves its ability to secure against them. Furthermore, the presented protocol provides reliable and cheaper NFTs than the classical one.
\end{abstract}


\section{Introduction}
The modern digital era is characterized by heightened trust and security concerns, especially regarding money. However, due to its digital ledger technology, blockchain is currently the most trusted technology for its ability to store continuously evolving data records in a decentralized and distributed network   \cite{nofer2017blockchain}, \cite{meng2018position}.
Blockchain utilizes a cryptographic hash function to encrypt communication between decentralized network nodes, immune to unauthorized attacks. The information stored in digital ledgers can be acquired and sold worldwide through cryptocurrency exchanges; however, the exchange of funds is recorded in a public ledger, and your coins are stored in your digital wallet \cite{zheng2018blockchain}. Blockchains are integrated innovations that consolidate computer technologies with infrastructure, network, proof of stake, data, and applications. Blockchain technology enables the creation of a transparent, distributed, cost-efficient, and adaptive environment where each transaction can be audited and audit trails can be created. 
With blockchain technology, trustability, auditability, immutability, identification, persistency, credibility, and transparency are all built-in, making it more effective \cite{monrat2019survey}.

Beyond cryptocurrency, blockchain technology has numerous applications in areas such as financial services \cite{treleaven2017blockchain} and social services, risk management \cite{fu2019big}, healthcare facilities \cite{mcghin2019blockchain}, the internet of things (IoT) \cite{novo2018blockchain} to public and social services \cite{huo2022comprehensive},
Among the significant blockchain-based innovations that impact intellectual property are non-fungible tokens (NFT). Non-fungible Tokens are a type of non-interchangeable digital asset which cannot be substituted for its equivalent one. Uniqueness, traceability, programmability, indivisibleness, and atomicity enable NFT to become more promising \cite{wang2021non, chohan2021non}. This allows them to be traced throughout an immutable digital ledger, resulting in a verifiable asset history. A government or any centralized entity would not manage these tokens but solely rely on peer-to-peer fashion among its users. Precisely NFTs act as a liaison between the digital and physical world through numerous applications in collectables, gaming,  virtual art \cite{trautman2021virtual, kugler2021non}, identity, private equity transactions, and real estate deals \cite{ mofokeng2018future}, ticketing events\cite{regner2019nfts}, intellectual property assets \cite{bamakan2022patents} to name a few.The NFT market has gone vertical; in just one year, the NFT market has taken substantial rise from total daily sales of about USD 183,121 in 2020 to an average of USD 38 million in 2021 \cite{pinto2022nft}\\

Classical cryptography is like some puzzle means security is ensured by complexity. The key to the security of classical computers is the limitation of the computational power of a classical computer. The emergence of powerful supercomputers and quantum computers jeopardised the security of classical encryption. We can solve these issues with the help of quantum computers, where the law of quantum physics ensures security.

 Notwithstanding the importance of the topic, few studies have been conducted on non-fungible tokens; however, to our knowledge, we are the first to incorporate it into quantum computing. A novel protocol for preparing a quantum non-fungible token has been developed. The NFT is mounted on a blockchain made using doubly hypergraph states. The information is added as the weight of the quantum state further, and the entanglement of weighted doubly hypergraph states replaces the classical cryptographic hash functions. The security of the proposed protocol is guarded by the law of quantum physics, making it safe from all types of classical and quantum attacks. Also, an experimental demonstration of the protocol using a cloud-based quantum computer IBM Q Casablanca established in Casablanca, the largest city in Morocco, and compared the experimental result with the expected result \cite{ibm}.\\
The contributions of this work are mentioned below:
\begin{itemize}
    \item An efficient protocol for preparing a quantum non-fungible token is proposed on a blockchain using doubly hypergraph states where entanglement took over the classical cryptographic hash functions.
    \item  A hybrid approach to making it more reliable and environmentally benign involves Proof of Stake.
    \item  The aforementioned protocol is tested and simulated using the IBM cloud-based quantum computer, and fidelity of 0.8 is achieved in the current accessibility.
    \item The designed protocol proves its performance, adequacy, and efficiency against various possible attacks.
\end{itemize}
Our paper is organized in the following manner, sec.\ref{sec:6} discusses the fundamental constituents of our QNFT protocol. sec.\ref{sec:2} represents the prototype design to develop quantum NFT. Security and adequacy of the scheme are examined in sec.\ref{sec:3}. Experimental realization of our protocol is carried out in sec.\ref{sec:4}. sec.\ref{sec:5} concludes the paper with possible future directions.
\section{Fundamentals of QNFT Protocol}\label{sec:6}
We put forth quantum NFT protocol, in which token is minted on the blockchain by employing doubly hypergraph states. Proof of stake is utilized to develop an environmentally friendly, productive protocol. We explain the primary constituents of our protocol.
\subsection{\label{sec:level1}Proof of Stake}
A stake characterizes the esteem or money one tends to wager on a particular outcome, and the strategy is named staking. Proof of stake \cite{kiayias2017ouroboros} (POS) may be a category of a mechanism utilized for blockchain. It works by choosing the validators in proportion to their stake within the associated cryptocurrency. We tend to append a blockchain group action to the blockchain itself, so it is frequently recognized. Validators perform this included factor to make it safely. We have proposed a mechanism to prevent a malicious user from being built up by imposing the compulsion that validators ought to have the number of blockchain tokens. It needs potential attackers to accumulate an outsized fraction of the tokens on the blockchain to mount an attack.\\
A validator stake is outlined by the product of the number of coins with the number of times a single user has controlled them. In layman's dialect, the individual will mine or validate block transactions agreeing to the number of coins they hold. This proposes that the more coins closely held by a miner, the more mining control they have. It is an alternate to Proof of Work (POW) \cite{gervais2016security}, the original consensus algorithm in blockchain technology, to make sure transactions and incorporate new blocks to the chain. For case, Peercoin, Nxt, Blackcoin, and ShadowCoin all work on proof of stake mechanism.\\
One of the significant advantages of Verification of Stake(POS) \cite{vasin2014blackcoin} is that they are energy efficient. Since all the nodes appear to be not competing against one another to associate a replacement block to the blockchain, energy is spared inside the proof of stake. Moreover, it is decentralized as rewards are proportional to the amount of stake. There is an issue of nothing at stake with this as there is no drawback to the nodes just in case they bolster numerous blockchains in the event of a blockchain split. Hence, each fork can cause multiple blockchains, and validators can work; additionally, the nodes inside the network can never reach a consensus. \\
To begin with, the operational execution of a proof-of-stake cryptocurrency was Peercoin \cite{xue2018proof}. There are sporadic proposals for Ethereum to alter from a POW to a POS mechanism.
\begin{figure}[]
    \centering
    \includegraphics[scale = .9]{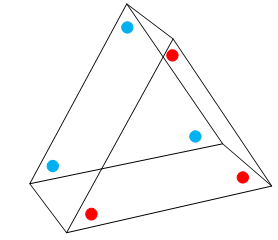}
    \caption{Representation of entanglement between six qubits double hypergraph state}
    \label{fig: 2}
\end{figure}
\subsection{Quantum double hypergraph states}
\begin{figure}[]
\centering
\includegraphics[scale = .6]{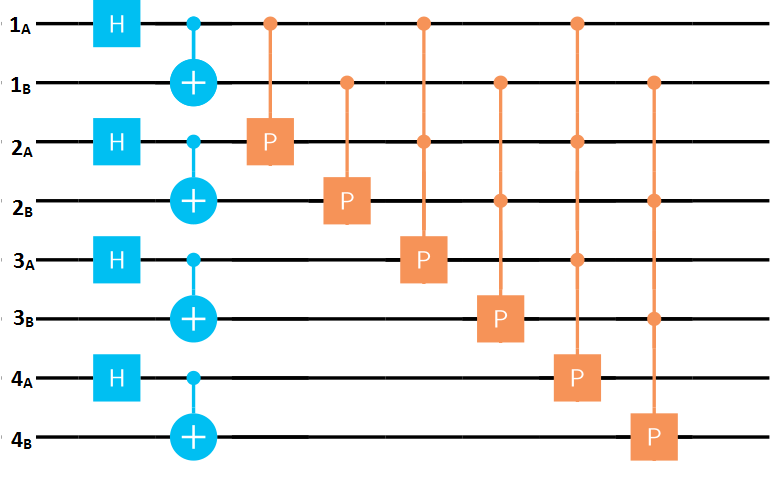}
\caption{Circuit to create quantum double hypergraph state.}
\label{fig .1}
\end{figure}

 Quantum hypergraph states are profoundly entangled multipartite quantum states based on a mathematical hypergraph. The quantum states reside on the vertices of the hypergraph, and the edges appear as attachments to other qubits, forming a non-separable many-body quantum state. The double hypergraph is comparative to the hypergraph state \cite{rossi2013quantum}, but each vertex have a parallel vertex shown in Fig. \ref{fig: 2}. We propose a methodology for creating a fundamentally quantum blockchain, using such states' entanglement as a replacement for traditional ledger and hash functions. A fundamental clarification of a hypergraph state is shown below. A comparable quantum state may be made from a mathematical hypergraph with k hyperedges (i.e., hyperedges connecting 2k qubits) and n vertices. The hypergraph's number of vertices breaks even with n, the double number of qubits within the quantum framework. All qubits are at first in a pair of bell states. One qubit of the bell state is class A and the second qubit is represented as class B. A controlled-phase operation with a phase angle of $\pi/2$ is then performed on each k-hyperedge of class A and B, shown in Fig. \ref{fig .1}. A double hypergraph with four vertices 1,2,3,4 where each vertex contains two qubits represented with classes A and B. Each qubit of class A and B are entangled in a bell state with each other at their respective vertices. The weighted double hypergraph states can be created by adding a phase to each qubit by local operation.

\section{Prototype design and development of Quantum NFT}

We have proposed a protocol for Quantum non-fungible tokens. Our protocol can potentially replace the classical NFT with the assistance of entanglement of a double-weighted hypergraph state, where the sell-off and statement of the victor can be chosen either classically or with quantum auction protocol. After announcing a victor, we create a token concurring with the agreement; a token has a few random angles. Each block of the blockchain contains two qubits (entangled in bell state) first qubit stores the information of the proprietor and its information, and the second qubit stores the token's data (a token is a few random angles). We utilize the "weights," i.e., the phase carried by the hyperedges of the double-weighted hypergraph state, to encode the classical information. In this protocol, we created a unique, non-fungible quantum token and mounted it on the blockchain \cite{banerjee2020quantum}. We do not grant any physical state or physical token to the owner; instead, we mount the owner's name or id and the asset's information on the blockchain. The owner is also part of that blockchain. Let us look at each step gradually.\\
\label{sec:2}
\subsection{Creation of block of NFT}
NFT does not contain any physical data but contains information about the owner and art \cite{ante2021non} (link of art or name of art ). In our protocol, we have taken a bell state such that the first qubit of the bell state contains information about the owner and art, whereas the second qubit stores information of token (token is some random phase). 
We consider the information as a string of binary which will have a decimal equivalent of p. Two qubits combine to make a peer state defined as $\ket{\psi} = \frac{\ket{00}+\ket{11}}{\surd2}$ and introduce the relative phase ( p-value ) of the system as,
\begin{equation}
    \ket{\psi_{1A,1B}} =S(p_{1A})\otimes S(p_{1B})= \frac{\ket{00}+e^{i(\theta_{1A} + \theta_{1B})}\ket{11}}{\surd2}.
\end{equation}
Where $\theta_{1A}$ and $\theta_{1B}$ $\in (0,\frac{\pi}{2})$ is a function of $P$, $f(P)$, any bijective function can be chosen and known to the particular peer who is part of the block, and $\sum_{i}\theta_{iA} + \theta_{iB}$ and  $ < \pi $ $\forall i $ Here, the number of the block added to the chain is represented by i. Now the state $\ket{\psi_{1A,1B}}$ carries information of owner and art in its first phase (1A) and token in the second phase (1B), this two set of qubits is the peer of the blockchain. There is a mutual agreement upon consensus between the peers.\\
Likewise, all the peers encode the classical information following the same function added by the first peer. The chosen function can be any bijective function only known to the particular peer in this blockchain.
 Let us look at the example if we consider our information input as 110 and function is $\theta_i$ = $\frac{1}{2^{i-1}}\theta_1$ here $\theta_1$ = $\frac{\pi}{4}$ then 110 can be written as $\frac{\pi}{4}$ + $\frac{\pi}{8}$ + $0$ . This is how we can encode the name of owner and info of art. The length of info must be fixed,in our case we fix three digit binary number whose value will be around 50 to 100 to ensure enough space for information. 

\subsection{Creation of Token}
\begin{figure}
\centering
\includegraphics[scale= .5]{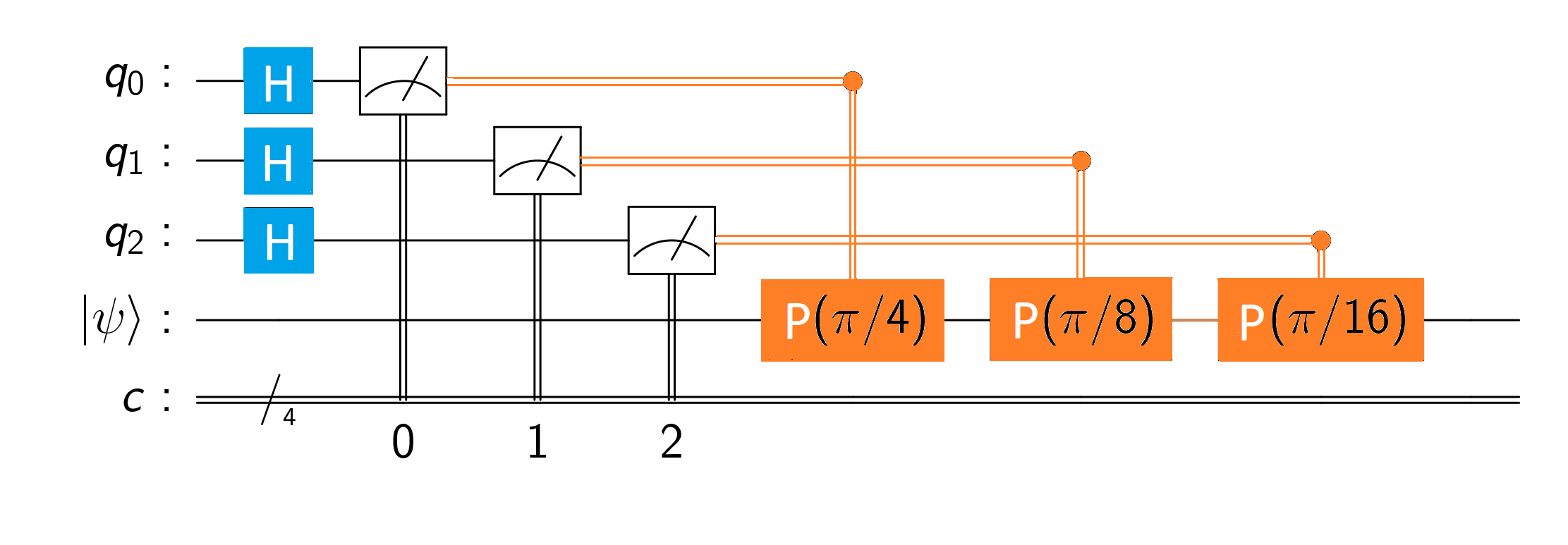}
\caption{Circuit to get random phase in $q_3$ where $q_0$,$q_1$ and $q_2$ are qubits for generation of random number.}
\label{fig 1}
\end{figure}
To create a token that is a random and unique phase, we use a Hadamard gate to place a qubit in a superposition of zero and one. After this, we measure that qubit. The measurement is purely random zero and one \cite{tamura2020quantum} (here, the randomness of this number depends on the laws of physics). Then, depending on the measurement outcome, we apply the phase gate. The angle of the phase gate is decided by a function $\theta_k$ $=$ $\frac{1}{2^{k+i}}\theta_1$, where i is the position of the binary number, and k is the number of peers. In Fig \ref{fig 1} we have to generate a random phase for the first peer using only three qubits where $\theta_1 = pi$. The number of Hadamard gates will decide the randomness of the phase. Hence the Hadamard gates must be substantial enough (around 20 to 50) to ensure randomness. Again this function takes care of uniqueness because the phase of the token depends on the number of blocks it belongs to on the chain. \\ 
\begin{figure*}
\centering
\includegraphics[scale = .7]{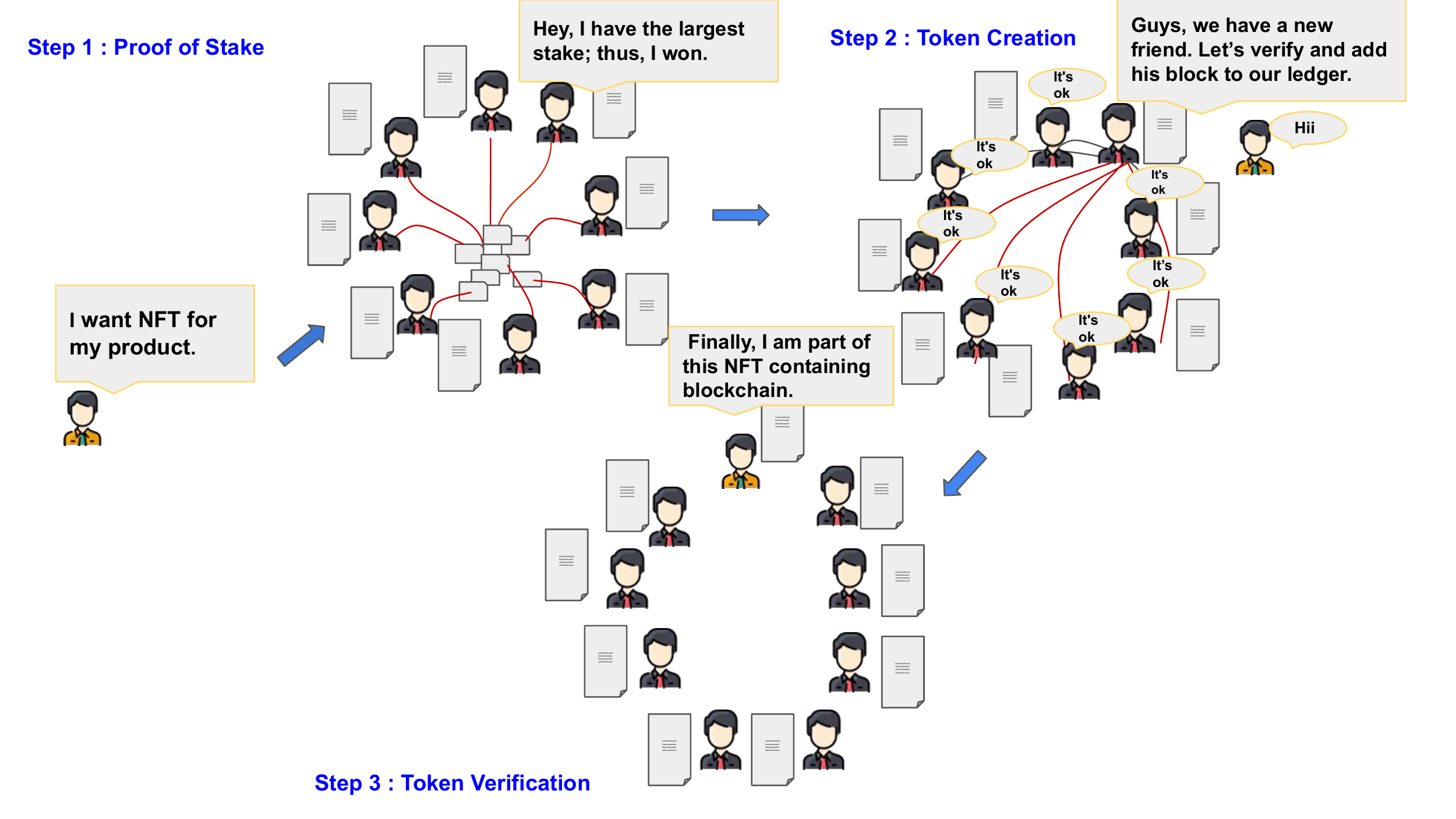}
\caption{ Elucidating our quantum NFT ecosystem for certain scenario such as addition of one peer to the main blockchain.}
\label{fig 6}
\end{figure*}

In equation \ref{table:1} to \ref{table:2}, we show that if we use only three Hadamard gates, then there are eight possible combinations, and for each binary string, we have different phases.\\

\begin{eqnarray}
\label{table:1}
\ket{000} = 0 + 0 + 0 = 0\\
\ket{001} = 0 + 0 + \frac{\pi}{4} = \frac{\pi}{4}\\
\ket{100} = \frac{\pi}{16} + 0 + 0 = \frac{\pi}{16}\\
\ket{110} = \frac{\pi}{16} + \frac{\pi}{8} + 0 = \frac{3\pi}{16}
\label{table:2}
\end{eqnarray}

This is how we get a relative random phase. As the number of Hadamard gate increases, the uniqueness of the phase will increase exponentially. If we use about 20 to 50 qubits, it is tough to guess the phase. This consensus will also ensure the randomness as well as the uniqueness of the token.\\
\begin{figure}
\centering
\includegraphics[width=\linewidth ]{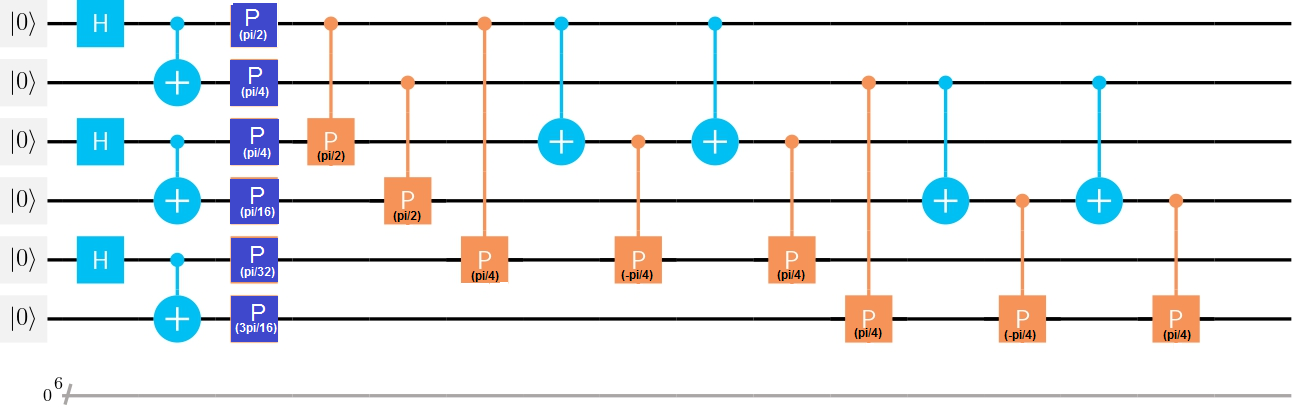}
\caption{Circuit for quantum 3-NFT ( $q_0$,$q_2$ and $q_4$ hold information of the owner and $q_1$,$q_3$ and $q_5 $ are tokens) prepared in IBM quantum experience(information encoded by applying suitable phase gate.}
\label{fig 2}
\end{figure}
\subsection{Verification of the blocks}
This step of verification is necessary to ensure that blocks are added according to consensus. The entanglement of double weighted hypergraph state allows us this step of verification, and every block should pass this verification step before adding it into our blockchain. According to our consensus, the peer who creates a token and mounted the owner name with asset sends the copy of the state to all peers and also informs them about relative phase $\theta_{mA}$ and $\theta_{mB}$.Now using QKD each peer verifies the state, 
\begin{equation}
    \ket{\psi_{mA,mB}} = \frac{\ket{00}+e^{i(\theta_{mA} + \theta_{mB})}\ket{11}}{\surd2}.
\end{equation}

Where the relative phase should be $\theta_{mA}$  = $(\frac{1}{n^{m-1}})\theta_{1A} $ and $\theta_{mB}$  = $(\frac{1}{n^{m-1}})\theta_{1B} $. He or she conveys a single copy of the state with each peer within the framework. When the peers get the qubit, they measure it on a basis $\ket{\pm_m}$ = $\frac{\ket{00} + e^{i(\theta_{mA}+\theta_{mB})}\ket{11}}{\sqrt{2}}$. In the event that the result of estimation is one, at that point they add the state in their native copy utilizing the $m − 1$controlled-P($\frac{\pi}{2}$)gate to each qubit of the block as appeared in Fig.\ref{fig 4}. The protocol is aborted for the other measurement results, and the peer is designated as untrustworthy as a result. and can be penalized in line with proof of stake.\\

\begin{figure}
\centering
\includegraphics[scale = .7]{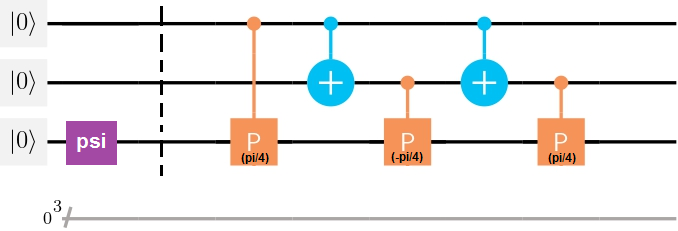}
\caption{Circuit for equivalent control- control phase gate .}
\label{fig 3}
\end{figure}

\begin{figure}
\centering
\includegraphics[scale = .6]{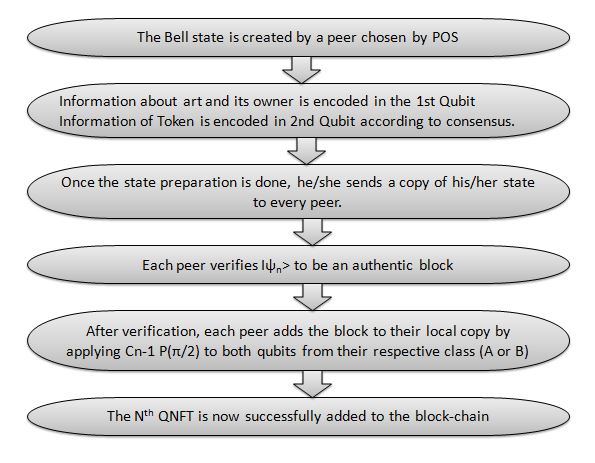}
\caption{Flowchart for creation of token and encoding it into blockchain.}
\label{fig 4}
\end{figure}

As we know that in IBM we don't have a control-control phase($\frac{\pi}{2}$) gate but we can apply some equivalent gate that shown in Fig \ref{fig 3}.

\begin{algorithm}
\caption{Quantum NFT}
\begin{algorithmic}[1]
\STATE All the NFT owners put their stake for POS process.
\STATE Probability of win in POS $\propto$ amount of stake. 
\STATE The winner create a state $\ket{\psi} = \frac{\ket{00}+e^{i(\theta_{mA} + \theta_{mB})}\ket{11}}{\surd2}.$

\STATE The state is sent to each peer for verification.
 \IF{ the state pass verification.}
\STATE The peer will add that state in their local blockchain copy.
\ELSE
\STATE The peer will discard the state.
\ENDIF
\end{algorithmic}
\end{algorithm}

\section{Security and Effectiveness}
\label{sec:3}
In our protocol, entanglement ensures the security of the blockchain in which QNFT is encoded. Entanglement deals with blockchain security only after the addition of a block in the chain by the peer. The authenticity of the blocks is verified by the process discussed previously. If the block does not succeed in the verification process, the corresponding peer who creates the block will lose the stake. The amount of stake is much higher than rewording, so the financial motivation will stop peers from doing unethical work.

Our convention depicts a blockchain as an entangled state, information is stored in phases, and n peers share n copy of the state at any given time. We do not have any public database; only $\theta_{p1}$ is shared with the assistance of QKD to all peers. In case we assume Eve altered with data, at that point, we can figure out the block that causes the collapse of the state and the peer whose state was annihilated. Once again, the particular state can be recouped as he/she knows the particular state without violating the no-cloning hypothesis.\\

In the following, we have considered a few possible attacks: intercept-and-resend, entangle-and-measure and man-in-the-middle in addition to classical threat. We will see how our protocol overcomes these attacks. 
\subsection{Intercept-and-resend attack}
We presume Eve as an eavesdropper who intercepts particles sent by Alice or any participant and resends a sequence of forged particles in the hope of succeeding the eavesdropping. In this case the forged particle can be identified in verification process when each peer measure the particle on a basis $\ket{\pm_m}$ = $\frac{\ket{00} + e^{i(\theta_{mA}+\theta_{mB})}\ket{11}}{\sqrt{2}}$. In that event, the result of estimation is one; at that point, they add the state in their native copy utilizing the $m − 1$ controlled-P($\frac{\pi}{2}$) gate to each qubit of the block as appeared in fig. 4. The protocol is aborted for the other measurement results, and the peer is labeled as untrustworthy as a result. He can be penalized in line with proof of stake.
\subsection{Entangle-and-measure attack}
\begin{figure}
\centering
\includegraphics[width = 7cm]{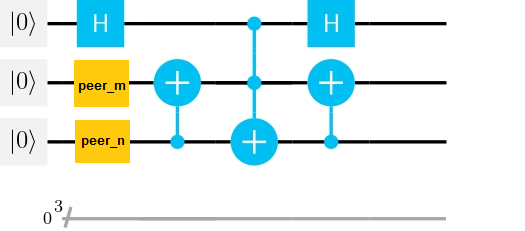}
\caption{Control swap test to test peer m and peer n is same or not.}
\label{fig 5}
\end{figure}

In this kind of attack, Eve uses a unitary operation $U_E$ to entangle an ancillary
particle on the transmitted quantum state and then measures the ancillary particle to
steal information.In this protocol, all information is stored in distributed phase, so even if Eve tries to entangle, he will not succeed.
\begin{eqnarray}
\begin{split}
        U_E\ket{\psi_{mA,mB}}&= \frac{a\ket{0}\ket{00}+b\ket{0}e^{i(\theta_{mA} + \theta_{mB})}\ket{11}}{\sqrt{2}}\\
    &+\frac{c\ket{1}\ket{00}+d\ket{1}e^{i(\theta_{mA} + \theta_{mB})}\ket{11}}{\sqrt{2}}
\end{split}
\\
\begin{split}
        U_E\ket{\psi_{mA,mB}} &= \frac{a\ket{+}\ket{00}+b\ket{+}e^{i(\theta_{mA} + \theta_{mB})}\ket{11}}{\sqrt{2}}\\
    &+\frac{c\ket{-}\ket{00}+d\ket{-}e^{i(\theta_{mA} + \theta_{mB})}\ket{11}}{\sqrt{2}}
\end{split}
\end{eqnarray}
Where $ e^{\theta_{mA}+\theta_{mB}} $ is combined phase of token and information of owner. Hence getting phase individual token is nearly impossible for invader if block is integrated in proposed blockchain. If Invader entangle with any block the controlled swap operation \cite{moulick2016quantum} is used to verify whether two states are the same or not. In this operation we use 3 qubits; $q_0$ is the test qubit , $q_1$, and $q_2$ are $peer_m$ and $peer_n$ respectively. If the test outcome is 0, then the test is successful; otherwise, it fails. With this controlled swap, we can compare two copies of the blockchain. This operation will not work for individual peers. A circuit for control swap test is shown in fig \ref{fig 5}.

\subsection{Man-in-the-middle attack}

Eve, the attacker, can use this type of attack to secretly relay and possibly manipulate communication between two parties who believe they are directly communicating. As a result, Eve can steal information from intercepted particles and also disrupt communication via re-sending forged particles. However, this type of attack cannot penetrate our protocol because our verification procedure protects it. 

We need to incorporate quantum identity authentication into our protocol, which can assist a party in determining whether another party is legitimate or not. When two parties interact, they must first validate each other's identities via quantum identity authentication, after which the attacker will be unable to impersonate a participant and connect with others. 
\begin{table*}
      
      \begin{tabular}{|p{4cm}|p{6cm}|p{6cm}|}
      \hline
       Threat   & Classical Security Issues & Quantum Solutions \\
      \hline
      \hline
      Spoofing (Authenticity)   &  Authentication vulnerabilities might be exploited by a hacker. A user's private key might be stolen.& Information is secure with entanglement and tempering with entanglement is not possible\\

     \hline
     Tampering   &  Manipulation with data store  & All the data shared between the peers using QKD\\
      
     (Integrity)   & outside the blockchain is possible.& is very secure\\
     \hline
      Repudiation   &  Binding of hash data with & No cloning theorem says  \\
      
     (Non-repudiability) & an attacker’s address possible. & Quantum state can not be clone\\
     \hline
     Information disclosure   & To link a particular NFT buyer or seller, an  &To exploit information attacker have to measure  \\
      
     (Confidentiality) & attacker can exploit the hash and transaction. &  the state which will cause collapse of state\\
     \hline
     
    \end{tabular}
    \caption{Security lapses with classical NFT and their quantum solutions}
    \label{tab: table 1}
\end{table*}
\subsection{Classical Threat}
The primary vulnerabilities associated with classical NFT are Spoofing, Tampering, Repudiation, and Information disclosure. Our protocol will attain a higher transparent system with quantum technology.\\

Spoofing is the capacity to imitate another entity on the same framework (for case, an individual or a computer), which correlates to genuineness. All data is stored in an entangled state (double hypergraph state), and imitating entanglement is unattainable.
\\
Tampering is defined as the malicious alteration of NFT data to compromise its integrity. Tampering with information is only possible before adding it to the blockchain, so all information is shared using QKD to avoid any attack.
\\
The term "repudiation" refers to a situation in which the originator of a statement is incapable of disputing it, which is linked to the security highlight of non-reputability \cite{zhou1996fair}. To accomplice this, a hostile attacker might steal the hash data, or the hash information could connect with the assailant's address. In our convention, data is put away in a quantum state, and the no-cloning theorem says that cloning a quantum state is impossible, so our NFT is secure from this sort of attack.
\\
When confidential information is exposed to unapproved users, usually known as information tampering, however, in our protocol, if an attacker wants to get information, he needs to apply a measurement operator that will lead to the collapse of the block and can effortlessly be taken note of by the owner.\\
 Table \ref{tab: table 1} displays how our protocol conquers the security lapses as to classical one.

\section{Discussion and evaluation}
\label{sec:4}
We will discuss in a manner our protocol fulfills all the basic properties of NFT. \\
\subsubsection{Token has unique identity}
A token is some random phase that builds using 200 to 300 qubits if we use only 20 qubits we have phase 1 in around 10 lakhs so it is extremely unlikely to have two tokens have the same phase even in 20 qubits so if we use such a large number it is nearly impossible that two tokens have the same phase, this is how we will ensure the uniqueness of token.
\subsubsection{Token must be non-fungible}
As the token is unique and the name of the owner and art is also attached to it due to this token can not be interchangeable. Non-fungibility does not mean one can not sell the art again rather it tells art is not interchangeable with each other, this is ensured by the uniqueness of token.
\subsubsection{Proof of ownership}
The token is minted on the blockchain and can be tracked to give the owner proof of ownership. In sec. \ref{sec:2} we addressed in detail how we have added token in our blockchain, One block consists of two qubits. The first qubit stores the information of the owner and its art, while the second qubit is reserved for tokens. According to our protocol, the owner is part of the blockchain as he/she is a peer of the blockchain. This gives the owner proof of ownership.
\subsubsection{Transparent operation.}
Each peer has their own copy of the blockchain. All the NFT operations including mounting, purchasing, and selling are open to the public. 
\subsubsection{Availability, Tamper-resistance and Usability}
As NFT is a distributed system, one or two fabrication does not affect the whole system. The system will always be available for the public to use. Entanglement ensures the whole system's security, so tempering it is not possible. Each block is instantaneously added to all blockchain copies, so each copy of the blockchain is up to date, which makes it user-friendly.
We have carried out our experiment on IBM Quantum Experience. Even though our protocol is planned for distributed ledger frameworks, we have executed our circuit on the IBM Quantum computer as a proof of concept. A set of circuits that work to complete our whole protocol. We have illustrated our blockchain circuit, a significant and principal portion of our entire protocol. This blockchain circuit is a 4-qubit system that incorporates a large set of quantum gates alongside our group of circuits. It contains two blocks of NFT. We run our circuit on 'IBM Q Casablanca,' a 7-qubit Quantum computer exclusively accessible to premium users. While testing, the most extreme number of shots, i.e., 8192, have been considered. We obtained a density matrix pretty much closer to the simulated matrix, which once more we get by running our circuit through 100000 shots on the QASM simulator. It gives us a density matrix that is nearly equal to the theoretical density matrix. We have calculated quantum state tomography to discover fidelity between the simulated density matrix and the experimental density matrix. The fidelity of 0.80 is achieved.\\

\subsection{Experimental realization}
\begin{figure}[]
\centering
\includegraphics[width=\linewidth]{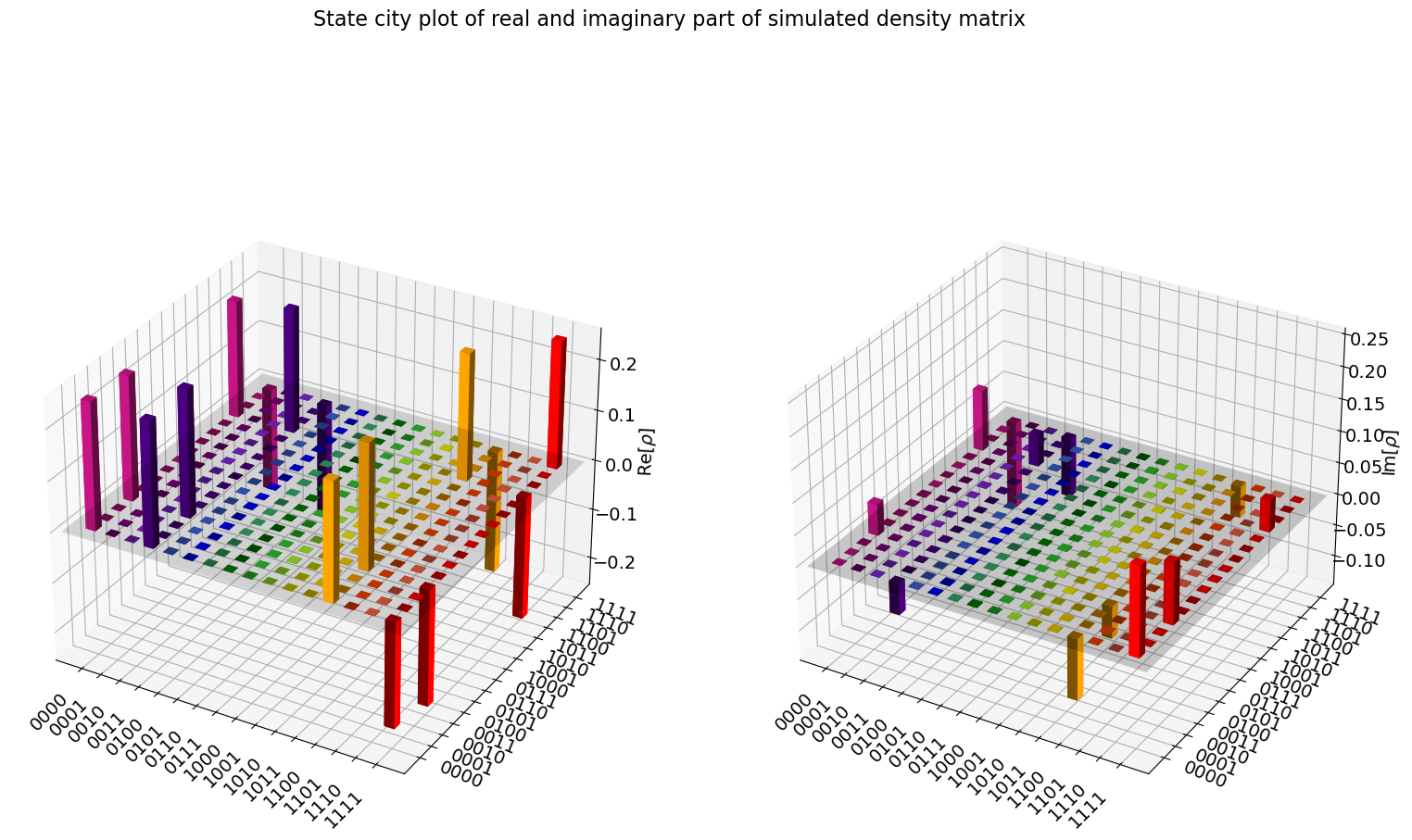}
\caption{City plots of real (left) and imaginary (right) parts of a simulated density matrix.}
\label{fig 9}
\end{figure}
\begin{figure}
    \centering
    %\begin{minipage}[b]{0.5\textwidth}
    \includegraphics[width= \linewidth]{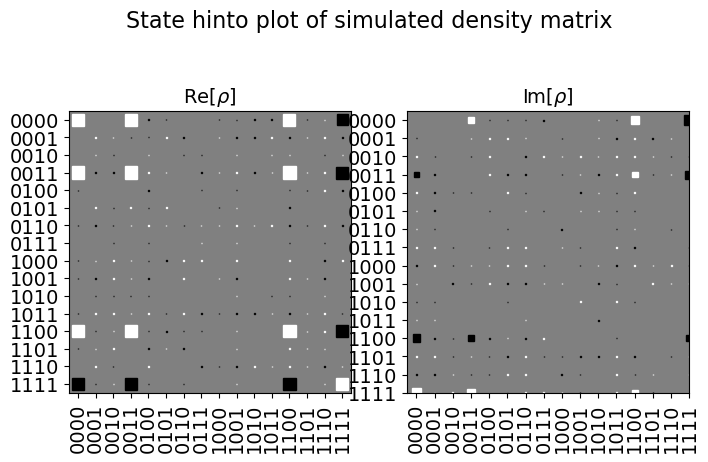}  
    \caption{Hinton plots of real (left) and imaginary (right) parts of a simulated density matrix.}
    \label{fig-11}
    %\end{minipage}%
    %\begin{minipage}[b]{0.5\textwidth}
    \includegraphics[width=\linewidth]{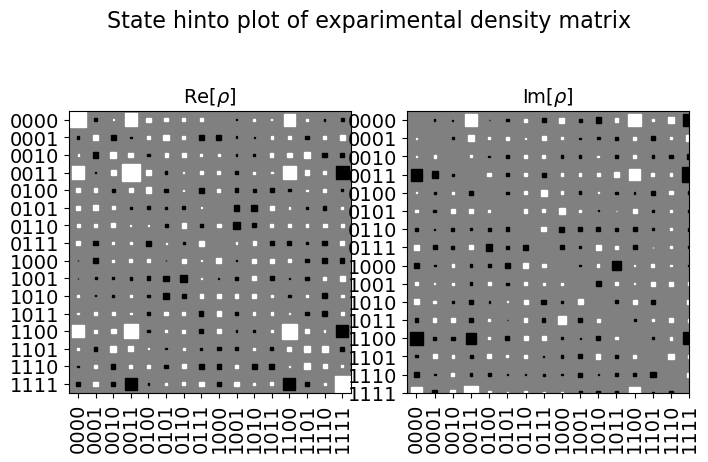} 
    \caption{Hinton plots of real (left) and imaginary (right) parts of a experimental density matrix.}
    \label{fig-12}
    %\end{minipage}
\end{figure}
\begin{figure}
\centering
\includegraphics[width=\linewidth]{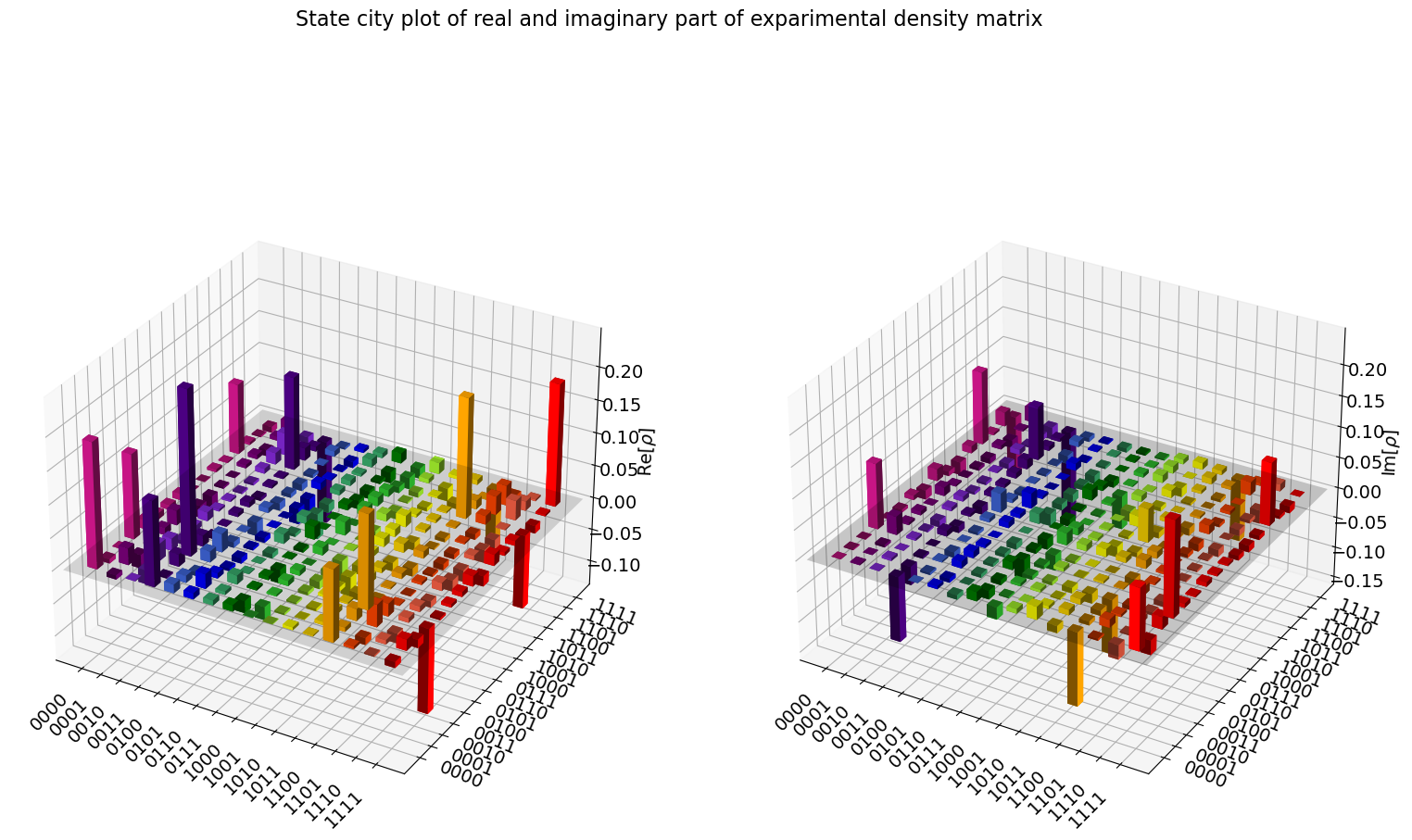}  
\caption{City plots of real (left) and imaginary (right) parts of a experimental density matrix.}
\label{fig-10}
\end{figure}

For our experiment, we have assumed the first block $q_0$, which holds the information about the owner and asset of the 1st nft, denoted as $\pi/16 $. The random token $q_1$ is also $\pi/16$, and the second block $q_2$ taken as $\pi/32$, which is carries particulars of the owner and asset of the 2nd nft, random token $q_3$ is also taken $\pi/32 $. All these values are according to the consensus. A city plot is the conventional representation of quantum states, in which the real and imaginary (image) components of the state matrix are represented like a city. Fig.\ref{fig 9} is the state city plot of the simulated density matrix, and Fig \ref{fig-10} is the state city plot of the experimental density matrix. The resemblance between the experimental and simulated (theoretical) density matrices can easily be seen, indicating that the effect of device noise is minimal.\\
Fig \ref{fig-11} and Fig \ref{fig-12} is the Hinton representation of simulated and experimental density matrix respectively. The element's size shows the value of the matrix element in a state Hinton plot, which is similar to a city plot.\\

We have also plotted a blockchain containing three QNFTs, shown in Fig \ref{fig 2}. For this plot, we assumed the phases as( $\pi/2$, $\pi/4$), ($\pi/4$,$\pi/16$)  and ($\pi/32$, $3\pi/16$) for the first, second, and third block respectively.  The first phase of each block is information about the owner and assets, and the second phase of each block is the token. In order to add those blocks to our blockchain, we have to apply the respective control phase ($\theta = \pi/2$) operation. For the second block, we have to apply the control phase operation directly available on IBM. However, in the third block, we have to apply the control-control phase ($\theta = \pi/2$) operation not directly available on IBM. So we use the equivalent circuit shown in Fig \ref{fig 3}. Following the same manner, we can add multiple blocks to this chain. Fig \ref{fig 6} portrays simple animation elaborating how a new block is added to the existing NFT blockchain. In step 1, all peers put their stakes, and the peer with the largest stake has the largest probability of being selected. In step 2, the winner from POS creates the quantum state representing the block and sends every peer for verification. And in the last step, the new peer is added to the blockchain as a new member and owns a complete blockchain set.

\section{Conclusion}
\label{sec:5}
To summarize, We have designed a protocol for preparing a quantum non-fungible token. Rather than giving the owner a physical quantum state representing NFT, we mounted it on a blockchain created utilizing doubly hypergraph states. The entanglement of the weighted double hypergraph state supplanted the conventional cryptographic hash functions. Our protocol incorporates proof of stake (POS) to develop an environmentally friendly, productive protocol which is eventually more effective than proof of work (POW). With POS, a peer is chosen; the same peer is responsible for making fair tokens and sends that token to each peer. After confirmation, each peer will include the new peer in their particular blockchain. With the help of animation and several figures, we have elaborated a simple process of forming QNFT in the blockchain. We have too tossed light on the adequacy and secureness of a few possible renowned attacks such as intercept- and-resend, entangle-and-measure, man-in-the-middle attack, and classical threat employing a quantum computer. We put up a quantum blockchain with two blocks on the IBM seven-qubit processor "IBM Q Casablanca" as a proof of concept, and the fidelity of 0.80 is achieved. A higher number of shots and a low noise device can increase fidelity. With the rapid development of quantum hardware, we are hopeful our protocol will achieve higher fidelity shortly. \\
The application of this protocol is exceptionally noteworthy. This protocol can replace classical NFT, and with the property of quantum physics, NFT proprietors will get a more secure stage for recognizing their assets. Eventually, NFTs pave the way to possibly digitizing the intellectual property rights and tokenizing the resources.

\section*{Acknowledgements\label{qnn_acknowledgements}}
S.S.P would like to thank IISER Kolkata for providing hospitality during the course of project work and QUEST(DST/ICPS/QuST/Theme-1/2019/2020-21/01) for financial support. T.D acknowledges financial support by Dept. of Science and Technology, India –INSPIRE  Fellowship (IF180118). We would also like to acknowledge the IBMQ Experience, through which all the experimentation has been carried out.We are thankful to Bikash K. Behera for his discussion prior to the start of the project.

\bibliographystyle{IEEEtran}
\bibliography{nft.bib}

\end{document}